\documentclass[a4paper,12pt]{article}
\usepackage[a4paper,text={16.8cm,22.4cm}]{geometry}
\usepackage{amsmath,amssymb,bm,psfrag,graphicx}
\RequirePackage[sort&compress,square,comma,numbers]{natbib}

\allowdisplaybreaks 
\begin{document}

\begin{titlepage}

\begin{flushright}
MZ-TH/12-51\\
December 11, 2012
\end{flushright}

\vspace{0.2cm}
\begin{center}
\Large\bf
Higgs-Boson Production at Small Transverse Momentum
\end{center}

\vspace{0.2cm}
\begin{center}
Thomas Becher$^a$, Matthias Neubert$^b$ and Daniel Wilhelm$^b$\\
\vspace{0.4cm}
{\sl 
${}^a$\,Albert Einstein Center for Fundamental Physics, Institut f\"ur Theoretische Physik, Universit\"at Bern, Sidlerstrasse 5, CH--3012 Bern, Switzerland\\[0.3cm]
${}^b$\,PRISMA Cluster of Excellence \& Mainz Institute for Theoretical Physics\\ 
Johannes Gutenberg University, 55099 Mainz, Germany}
\end{center}

\vspace{0.2cm}
\begin{abstract}
\vspace{0.2cm}
\noindent 
Using methods from effective field theory, we have recently developed a novel, systematic framework for the calculation of the cross sections for electroweak gauge-boson production at small and very small transverse momentum $q_T$, in which large logarithms of the scale ratio $m_V/q_T$ are resummed to all orders. This formalism is applied to the production of Higgs bosons in gluon fusion at the LHC. The production cross section receives logarithmically enhanced corrections from two sources: the running of the hard matching coefficient and the collinear factorization anomaly. The anomaly leads to the dynamical generation of a non-perturbative scale $q_*\sim m_H\,e^{-{\rm const}/\alpha_s(m_H)}\approx 8$\,GeV, which protects the process from receiving large long-distance hadronic contributions. We present numerical predictions for the transverse-momentum spectrum of Higgs bosons produced at the LHC, finding that it is quite insensitive to hadronic effects.
\end{abstract}
\vfil

\end{titlepage}

\section{Introduction}

The transverse-momentum spectra of electroweak bosons are among the most basic observables at hadron colliders. At large transverse momentum $q_T$ these spectra can be computed in fixed-order perturbation theory. On the other hand, if the transverse momentum is much smaller than the boson mass, higher-order corrections are enhanced by large Sudakov logarithms and must be resummed. This resummation was achieved a long time ago by Collins, Soper, and Sterman (CSS) \cite{Collins:1984kg} and has been implemented to high accuracy in several numerical codes \cite{Ladinsky:1993zn,Cao:2009md,deFlorian:2011xf,Bozzi:2010xn,Wang:2012xs}. In addition to the resummation at next-to-next-to-leading logarithmic (NNLL) accuracy, these programs include the matching to the next-to-next-to-leading (NNLO) fixed-order result \cite{Catani:2011kr}. For vector-boson production, we have revisited the resummation in the context of soft-collinear effective theory (SCET) \cite{Bauer:2000yr,Bauer:2001yt,Beneke:2002ph} and have derived an all-order factorization theorem for the cross section at small $q_T$ \cite{Becher:2010tm}. The factorization theorem for the differential cross section is affected by a collinear factorization anomaly, which generates an additional dependence on the hard momentum transfer in the low-energy theory. It was shown that in position space this extra dependence takes the form of a pure power of the gauge-boson mass. Relating our result to the traditional CSS formula then allowed us to determine the last missing ingredient needed for resummation to NNLL accuracy. The presence and all-order structure of these additional anomalous logarithms in the effective theory was confirmed by \cite{Chiu:2012ir} and \cite{GarciaEchevarria:2011rb}, but had been missed in earlier work on transverse-momentum resummation in SCET \cite{Gao:2005iu,Idilbi:2005er,Mantry:2009qz}. The work \cite{Chiu:2012ir} derives the anomalous logarithms using a renormalization-group framework, in which the evolution is performed in rapidity instead of virtuality. Their final result agrees with the factorization formula derived in \cite{Becher:2010tm}.

In the recent paper \cite{Becher:2011xn}, we have used our formalism to analyze the differential cross section $d\sigma/dq_T$ at very small transverse momentum, where it exhibits quite remarkable properties. In this region, the spectrum is genuinely non-perturbative but dominated by short-distance physics and therefore calculable. For a boson of mass $m_V$, long-distance effects are suppressed by a dynamically generated scale $q_*\sim m_V\,e^{-{\rm const}/\alpha_s(m_V)}$, which is close to 2\,GeV for the case of $Z$-boson production. While the underlying mechanism was identified a long time time ago \cite{Parisi:1979se}, our framework has allowed us to systematically compute corrections also in this region. In \cite{Becher:2011xn} we have performed a detailed phenomenological study of $Z$-production at the Tevatron and the LHC and have investigated the numerical impact of long-distance effects. 

The region of low transverse momentum is also important for the study of the Higgs boson and its properties, since the background is reduced when additional radiation is vetoed. In practice, this is done by imposing a jet veto. Several recent papers have considered resummation for the Higgs-boson cross section in the presence of such a cut \cite{Banfi:2012yh,Becher:2012qa,Tackmann:2012bt,Banfi:2012jm,Liu:2012sz}, and the resummation is now known to NNLL accuracy \cite{Becher:2012qa,Banfi:2012jm}. In the present paper, we extend the formalism of \cite{Becher:2010tm} to the transverse-momentum spectrum of the Higgs boson and perform the resummation of the spectrum to NNLL accuracy. The Higgs case was also considered in \cite{Chiu:2012ir}, where expressions for the spectrum were presented at NLL order. Much of the extension is straightforward and some of the necessary ingredients were already given in \cite{Becher:2010tm}, but there are a few interesting differences to the vector-boson case. First of all, while there is a single way to combine the spins of the incoming quarks to produce a vector boson, there are two ways to combine the gluon spins to produce a spin-zero state. As a consequence, the cross section is not just given by the product of two collinear functions as in the usual CSS formula, but a sum of two products of collinear functions describing the two production mechanisms, and the resummation formula must be modified accordingly \cite{Catani:2010pd}. We show that the collinear anomaly is the same for both structures, and that the dependence on the large scale $m_H$ therefore arises as an overall factor in position space. We then compute the collinear functions at one-loop order. The other feature, which distinguishes Higgs production from the vector-boson case, is that the infrared protection mechanism discussed above is much more efficient. The numerical value of $q_*\approx 8$\,GeV is significantly higher than in the $Z$-boson case, and we show that long-distance hadronic effects have almost no impact on the Higgs-boson spectrum. We have implemented our resummed results for Drell-Yan, $W$, $Z$, and Higgs production in a public code {\tt CuTe} \cite{CuTe} and give phenomenological predictions based on this program.

\section{Factorization and resummation}
\label{sec:fact}

We consider the cross section for the production of a Higgs boson with mass $m_H$ and transverse momentum $q_T=|q_\perp|$ in gluon fusion at the LHC. The derivation of the factorization formula for the cross section proceeds exactly as in the case of the Higgs-production cross section defined with a jet veto, which we have recently considered in \cite{Becher:2012qa}. Our analysis there has been performed at fixed $q_\perp$ and rapidity $y$ of the Higgs boson, and the integration over the boson phase-space was carried out at the end. We can thus immediately use the result for the factorized cross section obtained in \cite{Becher:2012qa}, which reads 
\begin{equation}\label{sig1}
\begin{aligned}
   d\sigma 
   &= \sigma_0(\mu)\,C_t^2(m_t^2,\mu) \left| C_S(-m_H^2,\mu) \right|^2\,\frac{m_H^2}{\tau s}\,
    dy\,\frac{d^2q_\perp}{(2\pi)^2} \int d^2x_\perp\,e^{-iq_\perp\cdot x_\perp} \\
   &\quad\times 2 {\cal B}_c^{\mu\nu}(\xi_1,x_\perp,\mu)\,
    {\cal B}_{\bar c\,\mu\nu}(\xi_2,x_\perp,\mu)\,
    {\cal S}(x_\perp,\mu) \,,
\end{aligned}
\end{equation}
where $\xi_{1,2}=\sqrt{\tau}\,e^{\pm y}$ and $\tau=(m_H^2+|q_\perp^2|)/s$. The Born-level cross section is
\begin{equation}
   \sigma_0(\mu) = \frac{m_H^2\,\alpha_s^2(\mu)}{72\pi (N_c^2-1) s v^2} \,,
\end{equation}
where $\sqrt{s}$ denotes the center-of-mass energy of the LHC and $v$ is the Higgs vacuum expectation value. The Wilson coefficient $C_t$ multiplies the effective $ggH$ operator obtained after integrating out the heavy top quark, while the hard matching coefficient $C_S$ arises when this operator is matched onto an effective two-gluon operator in SCET. Moreover, we have defined
\begin{eqnarray}\label{BSdef}
   {\cal B}_c^{\mu\nu}(\xi,x_\perp,\mu)
   &=& - \frac{\xi\,\bar n\cdot p}{2\pi} \int dt\,e^{-i\xi t\bar n\cdot p}\,
    \sum_{X_c}\,\langle P(p)|\,{\cal A}_{c\perp}^{\mu,a}(t\bar n+x_\perp)\,|X_c\rangle\,
    \langle X_c|\,{\cal A}_{c\perp}^{\nu,a}(0)\,|P(p)\rangle \,, \nonumber\\[-1mm]
   {\cal S}(x_\perp,\mu)
   &=& \frac{1}{N_c^2-1}\,\sum_{X_s}\,
    \langle\,0\,|\,\big( S_n^\dagger S_{\bar n} \big)^{ab}(x_\perp)\,|X_s\rangle\,
    \langle X_s|\,\big( S_{\bar n}^\dagger S_n \big)^{ba}(0)\,|0\rangle \,.
\end{eqnarray}
Here ${\cal A}_{c\perp}$ is the gauge-invariant effective gluon field of SCET, and $S_n$, $S_{\bar n}$ denote soft Wilson lines. The soft function ${\cal S}$ describes the physics of soft gluons emitted from the colliding beam particles. The function ${\cal B}_c^{\mu\nu}$ is the standard transverse parton distribution function, first introduced in \cite{Collins:1981uw}. It describes the structure of the jet of collinear particles inside one of the colliding protons (the one moving along the light-like direction $n^\mu$), which is probed at small transverse distance $x_\perp$. The corresponding function ${\cal B}_{\bar c}^{\mu\nu}$ for the second beam jet, consisting of anti-collinear particles (moving along $\bar n$), is given by the same formula with the replacements $\bar n\to n$ and $c\to\bar c$. In \cite{Becher:2012qa}, the sum over hadronic intermediate states was restricted by the jet veto, while in the present case the sums in (\ref{BSdef}) are completely inclusive.

In the context of SCET, proton matrix elements of collinear fields off the light cone are usually referred to as beam functions, a term introduced in \cite{Stewart:2009yx} for fully unintegrated PDFs, which also depend on the other light-cone coordinate $x_-=\bar n\cdot x$, in contrast to the matrix elements in (\ref{BSdef}). The two-loop renormalization of these functions was studied in \cite{Stewart:2010qs}, and the one-loop matching onto standard PDFs was calculated in \cite{Jain:2011iu}. The results of these two papers can, however, not be used in the present context, because in the limit $x_-\to 0$ light-cone singularities arise, which make the definition of the transverse PDFs in (\ref{BSdef}) subtle. In the context of SCET this was first discussed in \cite{Becher:2010tm}, and we will now explain it in more detail.

\subsection{Collinear anomaly}

The beam-jet functions $ {\cal B}_c^{\mu\nu}$, $ {\cal B}_{\bar c}^{\mu\nu}$ and the soft function ${\cal S}$ suffer from light-cone divergences, which are not regularized by the conventional dimensional regularization procedure. These singularities cancel in the product of the three functions, but in order to make the individual objects well-defined an additional regularization is required \cite{Becher:2010tm}. At the end of the calculation the regulator can be removed, but a non-trivial effect remains: The additional regularization breaks a rescaling symmetry exhibited at the classical level by the effective Lagrangian of SCET (i.e., the property that the collinear matrix elements are invariant under a rescaling of the anti-collinear momenta and vice versa), which is not restored in the limit where the regulator is removed because it is spoiled by quantum effects. This ``collinear factorization anomaly'' manifests itself through a dependence of the product of soft and beam functions on the Higgs mass -- the large scale in the problem. For the Drell-Yan process, the all-order form of this anomaly was derived in \cite{Becher:2010tm}, and it was shown that the dependence on the hard scale (the electroweak gauge-boson mass in this case) takes the form of a pure power in $x_\perp$ space. We will now extend the derivation to the case of Higgs production and show that the form of the anomaly as well as the corresponding anomalous dimensions are spin independent. 

The simplest way to introduce the additional regularization in such a way that gauge invariance and the factorization properties of the cross section are maintained is to regularize the phase-space integrals analytically by replacing \cite{Becher:2011dz}
\begin{equation}\label{regdef}
   \int\!d^dk\,\delta(k^2)\,\theta(k^0) 
   \to \int\!d^dk\,\delta(k^2)\,\theta(k^0) \left( \frac{\nu}{k_+} \right)^\alpha
\end{equation}
in the sum over intermediate states in (\ref{BSdef}). The scale $\nu$ is inserted to restore the canonical dimension of the integrals, in analogy to the scale $\mu$ of dimensional regularization. Instead of using a single light-cone component $k_+=k\cdot n$ in the regulator term, one could also use the sum $k_+ +k_-=2k_0$. This last form is similar to what is used in the ``rapidity regularization scheme'' proposed in \cite{Chiu:2012ir}. For pertubative computations, using a single light-cone component is simplest, since SCET Feynman diagrams typically already contain light-cone denominators involving $k_+$ or $k_-$. With the form (\ref{regdef}) of the regularization, one finds that the soft function is given by scaleless integrals, and thus ${\cal S}(x_\perp,\mu)=1$ to all orders in perturbation theory \cite{Becher:2010tm}. We will no longer write it explicitly in the rest of the paper. If one were to use the energy $k_0$ instead of $k_+$ in (\ref{regdef}), the soft function would be non-trivial, but can be absorbed into the beam-jet functions without loss of generality. 

The light-cone component $k_+$ of the anti-collinear particles moving along the $\bar{n}^\mu$ direction is large, $k_+ \sim m_H$. Expanding in the regulator $\alpha$, the dependence of the anti-collinear beam-jet function on the regulator scale $\nu$ thus takes the form $\ln(\nu/m_H)$. On the other hand, the $k_+$ component of the collinear partons is small, $k_+\sim q_T^2/m_H\sim 1/(x_T^2 m_H)$, where $q_T^2=-q_\perp^2$ and $x_T^2=-x_\perp^2$. In the collinear beam-jet function the dependence on the scale $\nu$ thus arises in the form $\ln(\nu x_T^2 m_H)$. The requirement that the physical cross section (\ref{sig1}) must be independent of the analytic regulator scale $\nu$ can then be expressed as
\begin{equation}\label{nuindep}
   \frac{d}{d\ln\nu}\,{\cal B}_c^{\mu\nu}\Big(\xi_1,x_\perp,\ln(\nu x_T^2 m_H),\mu\Big)\,
   {\cal B}_{\bar c}^{\rho\sigma}\Big(\xi_2,x_\perp,\ln\frac{\nu}{m_H},\mu\Big) = 0 \,.
\end{equation}
In the factorization theorem (\ref{sig1}) the Lorentz indices of the beam functions are contracted. The fact that $\nu$ independence also holds without contracting the indices follows by considering the factorization theorem for the production of a general color-neutral tensor field $h^{\mu\nu}$. The corresponding factorization theorem has the same structure as (\ref{sig1}), except that the hard matching coefficient $|C_S|^2$ would now depend on the Lorentz indices of the tensor fields in the initial and final states. Since the logarithms in (\ref{nuindep}) have different arguments, the cancellation of the $\nu$ dependence among the different factors imposes a non-trivial constraint on the $m_H$ dependence of the product. As explained in detail in \cite{Chiu:2007dg,Becher:2010tm}, the above equation implies that the dependence of the product of the two functions on $m_H$ must be power like. We can thus rewrite the product in the form
\begin{equation}
\begin{aligned}\label{refact}
   {\cal B}_c^{\mu\nu}\Big(\xi_1,x_\perp,\ln(\nu m_H x_T^2),\mu\Big)\,
    {\cal B}_{\bar c}^{\rho\sigma}\Big(\xi_2,x_\perp,\ln\frac{\nu}{m_H},\mu\Big) 
    & \\
   = \left( \frac{x_T^2 m_H^2}{b_0^2} \right)^{-F_{gg}(x_T^2,\mu)}  
   & B_{g}^{\mu\nu}(\xi_1,x_\perp,\mu)\,B_{g}^{\rho\sigma}(\xi_2,x_\perp,\mu) \,,
\end{aligned}
\end{equation}
with $b_0=2e^{-\gamma_E}$. The new beam-jet function $B_{g}^{\mu\nu}(\xi,x_\perp,\mu)$ and the anomaly exponent $F_{gg}(x_T^2,\mu)$ are independent of $m_H$.  

Having determined the form of the anomaly, we now derive the scale dependences of the function $B_{g}^{\mu\nu}(\xi_1,x_\perp,\mu)$ and the exponent $F_{gg}(x_T^2,\mu)$. Their anomalous dimensions can be inferred from the requirement that the cross section must be independent of the renormalization scale $\mu$, which implies that the $\mu$ dependence of the product of beam functions must cancel against that of the hard function $\sigma_0\,C_t^2\,|C_S|^2$ in (\ref{sig1}). This leads to the renormalization-group (RG) equations
\begin{equation}\label{Bevol}
\begin{aligned}
   \frac{dF_{gg}(x_T^2,\mu)}{d\ln\mu} 
   &= 2\Gamma_{\rm cusp}^A(\alpha_s) \,, \\
   \frac{d}{d\ln\mu}\,B^{\mu\nu}_{g}(\xi,x_\perp,\mu)
   &= \left[ \Gamma_{\rm cusp}^A(\alpha_s)\,\ln\frac{x_T^2\mu^2}{b_0^2} - 2\gamma^g(\alpha_s) \right]
    B^{\mu\nu}_{g}(\xi,x_\perp,\mu) \,,
\end{aligned}
\end{equation}
where $\Gamma_{\rm cusp}^A$ is the cusp anomalous dimension in the adjoint representation, and $\gamma^g$ is the anomalous dimension of the collinear gluon field as defined in \cite{Becher:2009qa}. The fact that each component of $B^{\mu\nu}_{g}(\xi,x_\perp,\mu)$ renormalizes in the same way follows after considering the production of a general tensor field $h^{\mu\nu}$ and using that the anomalous dimension of the hard function is spin-independent \cite{Becher:2009cu}. Three-loop expressions for $\Gamma_{\rm cusp}^A$ and $\gamma^g$ can be found in \cite{Becher:2009qa}.

Lorentz invariance implies that the renormalized beam-jet functions can be decomposed as 
\begin{equation}\label{Bmunudec}
   B_g^{\mu\nu}(\xi,x_\perp,\mu)
   = \frac{g_\perp^{\mu\nu}}{2}\,B_{g}^{(1)}(\xi,x_T^2,\mu) +
    \left( \frac{x_\perp^\mu x_\perp^\nu}{x_\perp^2} - \frac{g_\perp^{\mu\nu}}{2} \right)
    B_{g}^{(2)}(\xi,x_T^2,\mu) \,,
\end{equation}
where the coefficients only depend on the invariant $x_T^2=-x_\perp^2$. It follows that
\begin{equation}\label{entanglement}
  2 B_g^{\mu\nu}(\xi_1,x_\perp,\mu)\,B_{g\,\mu\nu}(\xi_2,x_\perp,\mu)
  = \sum_{n=1,2}\,B_g^{(n)}(\xi_1,x_T^2,\mu)\,B_g^{(n)}(\xi_2,x_T^2,\mu) \,.
\end{equation}
The presence of the two different tensor structures (\ref{Bmunudec}) contributing to the Higgs cross section was first pointed out in \cite{Catani:2010pd}. As a result, the Higgs-production cross section does not simply factor into a product of two beam-jet functions, but instead it involves the above sum of products, which arises because the spin-0 state of two gluons is an entangled state. Rewriting the cross section (\ref{sig1}) in terms of the new beam-jet functions, we obtain
\begin{equation}
\begin{aligned}\label{sig3}
   \frac{d^2\sigma}{dq_T^2\,dy} 
   &= \sigma_0(\mu)\,C_t^2(m_t^2,\mu) \left| C_S(-m_H^2,\mu) \right|^2\,
    \frac{1}{4\pi} \int d^2x_\perp\,e^{-iq_\perp\cdot x_\perp} 
    \left( \frac{x_T^2 m_H^2}{b_0^2} \right)^{-F_{gg}(x_T^2,\mu)} \\
   &\quad\times \Big[ 
    B_g^{(1)}(\xi_1,x_T^2,\mu)\,B_g^{(1)}(\xi_2,x_T^2,\mu)
    + B_g^{(2)}(\xi_1,x_T^2,\mu)\,B_g^{(2)}(\xi_2,x_T^2,\mu) \Big] \,,
\end{aligned}
\end{equation}
which is analogous to eq.~(18) of our paper \cite{Becher:2010tm} on the Drell-Yan cross section for the production of electroweak gauge bosons. In the above formula, the three disparate scales $m_t\gg m_H\gg x_T^{-1}$ appear in factorized form, and it will be possible to resum logarithms of their ratios by controlling the $\mu$ dependence of the various functions in (\ref{sig3}) using RG equations. After performing the Fourier integral the scales $m_H$ and $q_T$ get intertwined in a complicated way. At small $q_T$ this gives rise to interesting phenomena, such as a non-perturbative, but short-distance dominated dependence of the cross section on $q_T/m_H$ and $\alpha_s(q_T)$ \cite{Becher:2010tm,Becher:2011xn,Frixione:1998dw}. 

\subsection{Refactorization}

As long as the transverse displacement $x_T$ is much smaller than the scale of long-distance interactions in QCD ($x_T\ll\Lambda_{\rm QCD}^{-1}$), formula (\ref{sig3}) for the cross section  can be simplified further, by matching the beam functions $B_g^{(n)}$ onto ordinary parton distribution functions (PDFs), thereby computing their dependence on $x_T^2$ in perturbation theory. The relevant matching relations read \cite{Collins:1984kg,Collins:1981uw}
\begin{equation}\label{OPE}
   B_g^{(n)}(\xi,x_T^2,\mu)
   = \sum_{i=g,q,\bar q} \int_\xi^1\!\frac{dz}{z}\,I_{g\leftarrow i}^{(n)}(z,x_T^2,\mu)\,
    \phi_{i/P}(\xi/z,\mu) \,, 
\end{equation}
which is valid up to hadronic corrections suppressed by powers of $\Lambda_{\rm QCD}^2\,x_T^2$. The cross section can then be written in the final form
\begin{equation}\label{fact1}
\begin{aligned}
   \frac{d^2\sigma}{dq_T^2\,dy} 
   &= \sigma_0(\mu)\,C_t^2(m_t^2,\mu)
    \left| C_S(-m_H^2,\mu) \right|^2 \sum_{i,j=g,q,\bar q}\,  
    \int_{\xi_1}^1\!\frac{dz_1}{z_1} \int_{\xi_2}^1\!\frac{dz_2}{z_2} \\
   &\quad\times \bar C_{gg\leftarrow ij}(z_1,z_2,q_T^2,m_H^2,\mu)\,
    \phi_{i/P}(\xi_1/z_1,\mu)\,\phi_{j/P}(\xi_2/z_2,\mu) \,,
\end{aligned}
\end{equation}
where
\begin{equation}\label{Cdef}
\begin{aligned}
   \bar C_{gg\leftarrow ij}(z_1,z_2,q_T^2,m_H^2,\mu)
   &= \frac{1}{4\pi} \int\!d^2x_\perp\,e^{-iq_\perp\cdot x_\perp}
    \left( \frac{x_T^2 m_H^2}{b_0^2} \right)^{-F_{gg}(L_\perp,a_s)} \\
   &\quad\times \sum_{n=1,2}\,
    I_{g\leftarrow i}^{(n)}(z_1,L_\perp,a_s)\,I_{g\leftarrow j}^{(n)}(z_2,L_\perp,a_s) \,.
\end{aligned}
\end{equation}
With a slight abuse of notation, we have traded the variables $x_T^2$ and $\mu$ in the functions $F_{gg}$ and $I_{g\leftarrow i}^{(n)}$ for new variables
\begin{equation}\label{defs}
   L_\perp = \ln\frac{x_T^2\mu^2}{b_0^2} \,, \qquad
   a_s = \frac{\alpha_s(\mu)}{4\pi}
\end{equation}
without changing the names of these functions. This notation will be convenient for our discussion below, and it conforms with the notation used in \cite{Becher:2011xn}. Integrating the double differential cross section (\ref{fact1}) over rapidity, we find
\begin{equation}\label{fact2}
   \frac{d\sigma}{dq_T^2} 
   = \sigma_0(\mu)\,C_t^2(m_t^2,\mu) \left| C_S(-m_H^2,\mu) \right|^2 \sum_{i,j=g,q,\bar q}\, 
    \int_{\tau}^1\!\frac{dz}{z}\,\tilde C_{gg\leftarrow ij}\big(z,q_T^2,m_H^2,\mu\big)\,
    f\hspace{-1.8mm}f_{ij}(\tau/z,\mu) \,,
\end{equation}
where the parton luminosities and new kernel functions are defined as
\begin{equation}
\begin{aligned}
   f\hspace{-1.8mm}f_{ij}(z,\mu)
   &= \int_z^1\!\frac{du}{u}\,\phi_{i/N_1}(u,\mu)\,\phi_{j/N_2}(z/u,\mu) \,, \\  
   \tilde C_{gg\leftarrow ij}(z,q_T^2,m_H^2,\mu)
   &= \int_z^1\!\frac{du}{u}\,\bar C_{gg\leftarrow ij}(u,z/u,q_T^2,m_H^2,\mu) \,.
\end{aligned}
\end{equation}
The factorized cross sections (\ref{fact1}) and (\ref{fact2}) receive power corrections in the two small quantities $q_T^2/m_H^2$ and $\Lambda_{\rm QCD}^2\,x_T^2$, which will not be indicated explicitly in our equations. 

A dependence on the hard scale $m_H$ enters formula (\ref{fact1}) for the double-differential cross section in two places: via the hard matching coefficient $C_S$ and via an $x_T$-dependent power of $m_H$ under the Fourier integral in (\ref{Cdef}). The latter effect is due to the collinear factorization anomaly \cite{Becher:2010tm}. As long as $x_T^2\ll\Lambda_{\rm QCD}^{-2}$, the anomalous exponent $F_{gg}$ can be calculated in perturbation theory, and at least up to three-loop order it is related to the corresponding exponent $F_{q\bar q}$ appearing in the Drell-Yan case by the Casimir-scaling relation \cite{Becher:2010tm}
\begin{equation}
   \frac{F_{gg}(L_\perp,a_s)}{C_A} = \frac{F_{q\bar q}(L_\perp,a_s)}{C_F}
    + {\cal O}(\alpha_s^4) \,.
\end{equation}
Using the known expression for $F_{q\bar q}$, we then find
\begin{equation}\label{Fgg2}
   F_{gg}(L_\perp,\mu)
   = \Gamma_0^A \left[ a_s\,L_\perp + a_s^2 \left( \beta_0\,\frac{L_\perp^2}{2}
    + K L_\perp + d_2 \right) + \dots \right] ,
\end{equation}
where $\Gamma_0^A=4C_A$ and $\beta_0=\frac{11}{3}\,C_A-\frac43\,T_F n_f$ are the one-loop coefficients of the cusp anomalous dimension and $\beta$ function, and 
\begin{equation}
   K = \frac{\Gamma_1^A}{\Gamma_0^A} 
    = \left( \frac{67}{9} - \frac{\pi^2}{3} \right) C_A - \frac{20}{9}\,T_F n_f \,, \qquad
   d_2 = \frac{d_2^g}{\Gamma_0^A} 
    = \left( \frac{202}{27} - 7\zeta_3 \right) C_A - \frac{56}{27}\,T_F n_f
\end{equation}
contain the relevant two-loop information. Because of Casimir scaling, these coefficients take the same values as in the case of Drell-Yan production. 

At one-loop order, the kernel functions $I_{g\leftarrow i}^{(n)}(z,L_\perp,a_s)$ are given by
\begin{equation}\label{Ires}
\begin{aligned}
   I_{g\leftarrow i}^{(1)}(z,L_\perp,a_s) 
   &= \delta(1-z)\,\delta_{gi} \left[ 1 + a_s \left( \Gamma_0^A\,\frac{L_\perp^2}{4}
    - \gamma_0^g L_\perp \right) \right]
    + a_s \left[ - {\cal P}_{g\leftarrow i}^{(1)}(z)\,\frac{L_\perp}{2}
    + {\cal R}_{g\leftarrow i}(z) \right] , \\
   I_{g\leftarrow i}^{(2)}(z,L_\perp,a_s) 
   &= a_s\,{\cal R}'_{g\leftarrow i}(z) \,,
\end{aligned}
\end{equation}
where $\gamma_0^g=-\beta_0$, 
\begin{equation}
\begin{aligned}
   {\cal P}_{g\leftarrow g}^{(1)}(z) 
   &= 8C_A \left[ \frac{z}{\left(1-z\right)_+} + \frac{1-z}{z} + z(1-z) \right]
    + 2\beta_0\,\delta(1-z) \,, \\
   {\cal P}_{g\leftarrow q}^{(1)}(z) 
   &= 4C_F\,\frac{1+(1-z)^2}{z}
\end{aligned}
\end{equation}
are the one-loop DGLAP splitting functions, and the remainder functions ${\cal R}_{g\leftarrow i}(z)$ and ${\cal R}'_{g\leftarrow i}(z)$ are given by
\begin{equation}
\begin{aligned}
   {\cal R}_{g\leftarrow g}(z) 
   &= - C_A\,\frac{\pi^2}{6}\,\delta(1-z) \,, \qquad
   &{\cal R}_{g\leftarrow q}(z) 
   &= 2C_F\,z \,, \\
   {\cal R}'_{g\leftarrow g}(z) 
   &= - 4C_A\,\frac{1-z}{z} \,, \qquad
   &{\cal R}'_{g\leftarrow q}(z) 
   &= - 4C_F\,\frac{1-z}{z} \,.
\end{aligned}
\end{equation}
The expression for $I_{g\leftarrow i}^{(1)}$ was calculated in \cite{Becher:2012qa}, while the result for $I_{g\leftarrow i}^{(2)}$ is new. The one-loop functions $I_{g\leftarrow g}^{(i)}$ were also computed in Appendix~C of \cite{Chiu:2012ir}.\footnote{The expressions given there however contain several misprints \cite{private}.} In \cite{Becher:2010tm}, the SCET matching coefficients were related to the collinear functions in the traditional CSS formalism. Using these relations and the NNLO results presented in \cite{Catani:2011kr}, it would be possible to extract the matching functions $I_{g\leftarrow i}^{(1)}(z,L_\perp,a_s)$ at two-loop order.

\subsection{Resummation of Sudakov logarithms}

The resummation of large logarithms in the cross section (\ref{fact1}) is accomplished by evolving the hard matching coefficients $C_t$ and $C_S$ to a scale $\mu$ at which the kernel functions $\bar C_{gg\leftarrow ij}$ in (\ref{Cdef}) can be calculated using a controlled perturbative expansion. The solutions of the corresponding RG equations are discussed in detail in \cite{Becher:2012qa}. For our numerical work we use relations (A.2) and (A.4) from this paper. For the Drell-Yan case, the proper choice of the factorization scale $\mu$ has been discussed in \cite{Becher:2011xn}. Since the resulting expressions for Higgs production are completely analogous, we will not repeat details of the derivations here but rather summarize the main physical insights and quote the final expressions. The most naive choice would be to set $\mu\sim x_T^{-1}$ inside the Fourier integral in (\ref{Cdef}), in which case $L_\perp$ would be a small logarithm for any choice of $x_T$. There are several disadvantages to such a treatment. First, since $x_T$ is integrated over all possible values, there would be no clear meaning to the scale $\mu$ in terms of a characteristic scale of the process. Second, setting the scale under the integral means that the integration unavoidably hits the Landau pole of the running coupling, giving rise to ambiguities in the numerical results. In the spirit of effective field theory, the scale $\mu$ should correspond to a physical scale in the underlying factorization theorem. We will choose it in such a way that {\em on average\/} the $x_T$-dependent logarithm $L_\perp$ is small, and denote the corresponding value by $\mu\sim\langle x_T^{-1}\rangle$.

Naively, one would expect that the transverse momentum $q_T$ and average transverse separation $\langle x_T\rangle$ are conjugate variables satisfying $q_T\sim\langle x_T^{-1}\rangle$. While this is sometimes true, the general situation turns out to be more complicated. After integration over $x_\perp$, the factorized dependence on $m_H$ and $q_T$ in (\ref{Cdef}) gets intertwined in a complicated way, and this gives rise to the peculiar effect that the two scales $q_T$ and $\mu\sim\langle x_T^{-1}\rangle$ decouple for very small $q_T$ \cite{Becher:2011xn}. When this happens depends on the value of the coefficient
\begin{equation}\label{etadef}
   \eta \equiv \Gamma_0^A\,a_s\ln\frac{m_H^2}{\mu^2}
   = \frac{C_A\alpha_s(\mu)}{\pi}\,\ln\frac{m_H^2}{\mu^2} \,.
\end{equation}
As long as $0<\eta<1$, one indeed finds that $\mu\sim\langle x_T^{-1}\rangle\sim q_T$, because contributions from large values $x_T\gg q_T^{-1}$ are suppressed due to the rapid oscillations of the phase factor of the Fourier integral, while contributions from small values $x_T\ll q_T^{-1}$ are phase-space suppressed. The situation changes, however, at very small transverse momentum, where with the prescription $\mu\sim q_T$ the value of $\eta$ reaches 1. We denote by $q_*$ the value of $\mu$ where this happens, i.e.\
\begin{equation}\label{qstar}
   q_* = m_H\exp\left( - \frac{2\pi}{\Gamma_0^A\alpha_s(q_*)} \right) 
   \approx m_H\exp\left( - \frac{2\pi}{\left(\Gamma_0^A+\beta_0\right)\alpha_s(m_H)} \right) ,
\end{equation}
where in the last step we have used the one-loop approximation for the running coupling. As long as $q_*$ is in the perturbative domain, one finds that at this scale $\langle x_T^{-1}\rangle$ decouples from $q_T$, and it remains a short-distance scale even in the extreme case where $q_T$ is taken to 0 \cite{Becher:2011xn}. Changing variables from $x_T$ to $L_\perp$ in the Fourier integral, one observes that the integrand exhibits a Gaussian peak with a width proportional to $1/\sqrt{a_s}$. The condition that at the peak the logarithm $L_\perp={\cal O}(1)$ implies that $1-\eta={\cal O}(a_s)$, indicating that the factorization scale must be chosen in the vicinity of $q_*$. We thus conclude that the proper scale choice is
\begin{equation}
   \mu\sim\langle x_T^{-1}\rangle\sim \max( q_T, q_* ) \,.
\end{equation}
In our numerical work below, we will use $\mu=q_T+q_*$ as the default choice for the factorization scale. Solving the first equation in (\ref{qstar}) numerically, we obtain $q_*\approx 7.7$\,GeV for $m_H=125$\,GeV, which is a short-distance scale well inside the perturbative domain. Due to the difference in color factors, this scale is significantly larger than in the case of the Drell-Yan production of electroweak gauge bosons, for which $q_*\approx 1.75$\,GeV \cite{Becher:2011xn}.

It follows from these arguments that the transverse-momentum distribution of Higgs bosons is protected from long-distance physics even for arbitrarily small $q_T$ -- a fact that in the context of the Drell-Yan process has been pointed out first a very long time ago in \cite{Parisi:1979se}. The resummed perturbative series for the cross section generates the scale $q_*$ dynamically, and even though this is a short-distance scale, it is related to the boson mass $m_H$ in a genuinely non-perturbative way. The scale $q_*$ also sets the magnitude of hadronic long-distance corrections, which turn out to be power-suppressed in the ratio $\Lambda_{\rm QCD}/q_*$.  The dynamical origin of this suppression was studied in detail in \cite{Becher:2011xn}. 
 We expect that these corrections are significantly smaller for Higgs production than for the Drell-Yan production of $Z$ and $W$ bosons. This expectation will be confirmed by our numerical studies presented below. 

The above discussion shows that we must distinguish two regions of transverse momenta. For $q_T\gg q_*$, the scale choice $\mu\sim q_T$ prevents that the logarithms $L_\perp$ give rise to large perturbative corrections. It is then consistent to count these logarithms as $L_\perp\sim 1$ and construct the perturbative series as a series in powers of $a_s$. A different situation is encountered for $q_T\ll q_*$. Even though the scale choice $\mu\sim q_*$ ensures that $L_\perp={\cal O}(1)$ on average, the Gaussian weight factor allows for significant contributions to the Fourier integral over a range of larger $L_\perp$ values with a width proportional to $1/\sqrt{a_s}$. It is then necessary to reorganize the perturbative expansion by adopting the modified power counting $L_\perp\sim 1/\sqrt{a_s}$. This implies that single-logarithmic terms $\left(a_s L_\perp\right)^n\sim a_s^{n/2}$ are always suppressed, whereas double-logarithmic terms $\left(a_s L_\perp^2\right)^n\sim 1$ are unsuppressed and must be resummed to all orders. To keep track of this fact, we introduce an auxiliary expansion parameter $\epsilon$ (which at the end is set to~1) and assign the power counting $a_s\sim\epsilon$ and $L_\perp\sim\epsilon^{-1/2}$. The terms contributing up to ${\cal O}(\epsilon)$ to the cross section have been derived in \cite{Becher:2011xn} using recursive solutions of the relevant RG equations. Adapting the resulting expression to the present case, we find that the hard-scattering kernels defined in (\ref{Cdef}) can be written in the form
\begin{equation}\label{Cfinal}
\begin{aligned}
   \bar C_{gg\leftarrow ij}(z_1,z_2,q_T^2,m_H^2,\mu)
   &= \frac12 \int_0^\infty\!dx_T\,x_T\,J_0(x_T q_T)\,\exp\big[ g_A(\eta,L_\perp,a_s) \big] \\
   &\quad\times \sum_{n=1,2}\,\bar I_{g\leftarrow i}^{(n)}(z_1,L_\perp,a_s)\,
    \bar I_{g\leftarrow j}^{(n)}(z_2,L_\perp,a_s) \,,
\end{aligned}
\end{equation}
where 
\begin{eqnarray}\label{gdef}
   g_A(\eta,L_\perp,a_s) 
   &=& - \big[ \,\eta L_\perp \, \big]_{\epsilon^{-1/2}}
    - \left[ a_s \left( \Gamma_0^A + \eta\beta_0 \right) \frac{L_\perp^2}{2}
    \right]_{\epsilon^0} \nonumber\\
   &&\mbox{}- \left[ a_s \left( 2\gamma_0^g + \eta K \right) L_\perp
    + a_s^2 \left( \Gamma_0^A + \eta\beta_0 \right) \beta_0\,\frac{L_\perp^3}{3} 
    \right]_{\epsilon^{1/2}} \\
   &&\mbox{}- \left[ a_s\,\eta d_2
    + a_s^2 \Big( K\Gamma_0^A + 2\gamma_0^g\beta_0 + \eta\,\big( \beta_1 + 2K\beta_0 \big) 
    \Big) \frac{L_\perp^2}{2}
    + a_s^3 \left( \Gamma_0^A + \eta\beta_0 \right) \beta_0^2\,\frac{L_\perp^4}{4} 
    \right]_{\epsilon} \nonumber\\
   &&\mbox{}- {\cal O}(\epsilon^{3/2}) \,. \nonumber
\end{eqnarray}
Note that we treat $\ln(m_H^2/\mu^2)$ as a large logarithm and count $\eta$ defined in (\ref{etadef}) as an ${\cal O}(1)$ variable. The auxiliary parameter $\epsilon$ counts the order in $a_s$ resulting (for $q_T\ll q_*$) after the $x_T$ integral in (\ref{Cfinal}) has been performed. The two terms given in the first line are unsuppressed and must be kept in the exponent of the integrand in (\ref{Cfinal}), whereas the remaining terms can be expanded in powers of $\epsilon^{1/2}$. It is important that the expansion is truncated at an integer power of $\epsilon$. The resulting integrals over the Bessel function can readily be evaluated numerically. 

We finally give the expressions for the collinear kernel functions corresponding to our modified power counting. We find
\begin{equation}\label{Imod}
\begin{aligned}
   \bar I_{g\leftarrow i}^{(1)}(z,L_\perp,a_s) 
   &= \delta(1-z)\,\delta_{gi} - \left[ a_s\,{\cal P}_{g\leftarrow i}^{(1)}(z)\,
    \frac{L_\perp}{2} \right]_{\epsilon^{1/2}} \\
   &\quad\mbox{}+ \left[ a_s\,{\cal R}_{g\leftarrow i}(z) 
    + a_s^2 \left( {\cal D}_{g\leftarrow i}(z) - 2\beta_0\,{\cal P}_{g\leftarrow i}^{(1)}(z)
    \right) \frac{L_\perp^2}{8} \right]_\epsilon + {\cal O}(\epsilon^{3/2}) \,,
\end{aligned}
\end{equation}
while $\bar I_{g\leftarrow i}^{(2)}$ coincides with $I_{g\leftarrow i}^{(2)}$ in (\ref{Ires}) up to higher-order terms in $\epsilon$. The corresponding contribution to (\ref{Cfinal}) is of ${\cal O}(\epsilon^2)$ and can be neglected to the order we are working. The quantities 
\begin{equation}\label{Ddef}
   {\cal D}_{g\leftarrow i}(z) 
   = \sum_{j=g,q,\bar q} \int_z^1\!\frac{du}{u}\,{\cal P}_{g\leftarrow j}^{(1)}(u)\,
    {\cal P}_{j\leftarrow i}^{(1)}(z/u)
\end{equation}
involve the convolutions of two DGLAP splitting functions. Following \cite{Becher:2011xn}, we find
\begin{equation}
\begin{aligned}
   &{\cal D}_{g\leftarrow g}(z) - 2\beta_0\,{\cal P}_{g\leftarrow g}^{(1)}(z) \\
   &= 64 C_A^2\,\Bigg[ \! \left( \frac{\ln\frac{(1-z)^2}{z}}{1-z} \right)_+ \!
    + \frac{1-2z+z^2-z^3}{z}\,\ln\frac{(1-z)^2}{z}
    - 2(1+z) \ln z + 3(1-z) - \frac{11}{3}\,\frac{1-z^3}{z} \Bigg] \\
   &\quad\mbox{}+ 16 C_A\beta_0 \left[ \frac{z}{\left(1-z\right)_+} + \frac{1-z}{z} + z(1-z) \right]
    + 32 C_F T_F n_f \left[ 2(1+z) \ln z + 1-z + \frac43\,\frac{1-z^3}{z} \right] , \\[2mm]
   &{\cal D}_{g\leftarrow q}(z) - 2\beta_0\,{\cal P}_{g\leftarrow q}^{(1)}(z) \\
   &= 16 C_A C_F \left[ \frac{1+(1-z)^2}{z}\,\ln\frac{(1-z)^2}{z} - \frac{2+6z+3z^2}{z}\,\ln z 
    - (1-z) \left( \frac{31}{3z} + \frac73 + \frac{4z}{3} \right) \right] \\
   &\quad\mbox{}+ 16 C_F^2 \left[ \frac{1+(1-z)^2}{z}\,\ln\frac{(1-z)^2}{z} + \frac{2\ln z}{z} 
    + 2 - \frac{z}{2} \right] . 
\end{aligned}
\end{equation}

Equations (\ref{Cfinal})--(\ref{Imod}) are our main results. With the help of these expressions, large logarithms can be resummed at NNLL order for arbitrarily small transverse momenta. For larger $q_T$ values, the additional terms contained in (\ref{gdef}) and (\ref{Imod}) reduce to higher-order terms proportional to $a_s^2$ and $a_s^3$, which can be kept without doing any harm. Our formula thus provides a smooth interpolation between the regions of small and very small $q_T$. In fact, it has been shown in \cite{Becher:2011xn} that the additional terms needed at very low $q_T$ serve an important purpose also at $q_T>q_*$, as they resum an asymptotically divergent, but Borel-summable subset of higher-order corrections in~$a_s$.

\begin{figure}[t!]
\begin{center}
\begin{tabular}{ccc}
\includegraphics[width=0.45\textwidth]{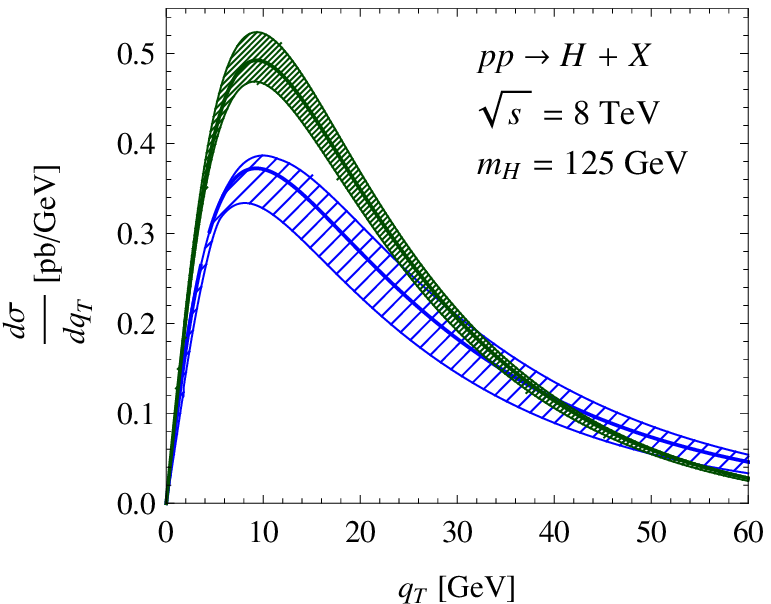} & &\includegraphics[width=0.465\textwidth]{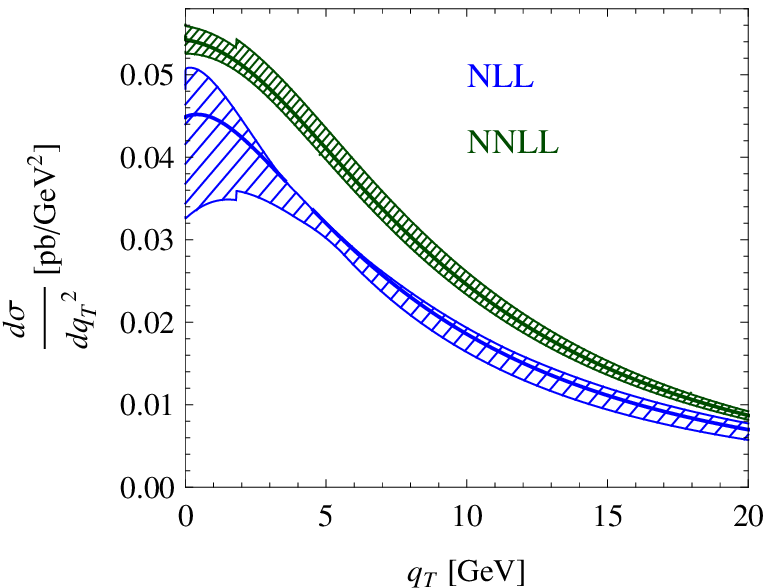} 
\end{tabular}
\end{center}
\vspace{-0.5cm}
\caption{\label{fig:NLLvsNNLL}
Resummed predictions for the transverse-momentum distribution of Higgs bosons produced at the LHC at NLL (blue bands) and NNLL order (green bands). The factorization scale $\mu$ is varied by a factor 2 about its default value $\mu=q_T+q_*$. The thick lines refer to the default scale choice. The discontinuities at $q_T=1.8$\,GeV arise from the change from 5 to 4 light flavors occurring when $\mu\le\frac12(q_T+q_*)=4.75$\,GeV.}
\end{figure}

In Figure~\ref{fig:NLLvsNNLL}, we present predictions for the resummed transverse-momentum distributions of Higgs bosons with mass $m_H=125$\,GeV produced in gluon fusion at the LHC. The blue bands correspond to the results obtained at NLL order, in which case we keep terms up to ${\cal O}(\epsilon^0)$ in (\ref{gdef}) and (\ref{Imod}) and use leading-order (LO) expressions for the Wilson coefficients $C_t$ and $C_S$ in RG-improved perturbation theory. The green bands show results with NNLL accuracy, which are obtained by retaining all terms shown in (\ref{gdef}) and (\ref{Imod}) and using next-to-leading-order (NLO) expressions for $C_t$ and $C_S$. We use $v=246.675$\,GeV for the Higgs vacuum expectation value and work with MSTW2008NNLO PDFs with $\alpha_s(M_Z)=0.1171$ \cite{Martin:2009iq}. In our numerical results, we include finite top-quark mass effects by working with the exact leading-order cross section, which can be found e.g.\ in \cite{Ahrens:2008nc}. Numerically, using $m_t=172.6$\,GeV, this leads to a 6.6\% increase in the cross section compared to the $m_t\to\infty$ limit. We do not include the $b$-quark contribution and electroweak effects, which are of similar size but opposite sign and tend to cancel each other. The combined effect of these would be a reduction of the cross section by about 1.5\%. The left plot in the figure shows the spectrum $d\sigma/dq_T$, while the plot on the right shows the distribution $d\sigma/dq_T^2$, which tends to a constant at the origin. Note that at NNLL order the value of the intercept is predicted with very good accuracy. In \cite{Becher:2011xn}, we have presented an explicit formula for the intercept for the case of the Drell-Yan process.

To estimate the theoretical uncertainty of our predictions, we vary the factorization scale by a factor 2 about the default value $\mu=q_T+q_*$. Our results also depend on the two hard matching scales $\mu_t$ and $\mu_h$. At the scale $\mu_t$ the top quark is integrated out, giving rise to an effective $ggH$ vertex. The associated matching corrections are small, and varying $\mu_t$ in the NNLL order cross section by a factor 2 about the default value $\mu_t=m_t$ changes the result by less than 1\%. The scale $\mu_h$ is associated with the hard momentum transfer in the $ggH$ coupling, as described by the time-like gluon form factor. In \cite{Ahrens:2008qu,Ahrens:2008nc}, we have shown that it is advantageous to evaluate the relevant matching coefficient $C_S(-m_H^2,\mu_h)$ with a time-like scale choice $\mu_h^2=-m_H^2$. This eliminates the large perturbative corrections arising when the gluon form factor is continued from space-like to time-like kinematics. When this is done, also the corrections to $C_S(-m_H^2,\mu_h)$ are of moderate size, and the effect of a variation of $\mu_h$ by a factor 2 on the cross section is again below 1\%. Since the variation of the factorization scale $\mu$ leads to the largest scale uncertainties by far, we use it to generate the error bands in the plots, keeping the hard matching scales $\mu_t$ and $\mu_h$ fixed at their default values.  The NNLL corrections have a noticable effect and strongly enhance the cross section in the peak region. From the right plot, we observe that the scale variation at NLL order is very small in the vicinity of $q_T=5\,{\rm GeV}$, because the predictions with high and low $\mu$ values cross each other. Near such a band crossing, the scale variation underestimates the theoretical uncertainty, and it is therefore not too surprising that the NLL and NNLL bands do not overlap in the low-$q_T$ region. Since  the prediction at NNLL order does not exhibit a band crossing, we believe that its scale variation provides a more reliable estimate of the theoretical uncertainty. Since we observe that the one-loop correction arising at NNLL order is larger than the NLL scale dependence suggests, we will be conservative when performing the matching to the fixed-order result in Section~\ref{sec:LHC} and adopt a matching scheme which yields larger scale uncertainties for the combined result than for the resummed result itself (see Figure~\ref{fig:matching_final}).

We proceed to study the impact on long-distance hadronic effects on the transverse-momentum distribution, which for the case of Drell-Yan production of electroweak bosons are known to have a non-negligible impact \cite{Ladinsky:1993zn,deFlorian:2011xf,Konychev:2005iy}. Following \cite{Becher:2011xn}, we model these effects by noting that the beam-jet functions $B_g^{(n)}(\xi,x_T^2,\mu)$ in (\ref{sig3}), which are nothing but transverse-position dependent PDFs, must vanish rapidly when the two gluon fields are separated by a transverse distance $x_T$ larger than the proton size. This motivates an ansatz of the form
\begin{equation}\label{Binonpert}
   B_g^{(n)}(\xi,x_T^2,\mu)
   = f_{\rm hadr}(x_T\Lambda_{\rm NP})\,B_g^{(n)\,\rm pert}(\xi,x_T^2,\mu) \,,
\end{equation}
where the perturbative functions $B_g^{(n)\,\rm pert}$ carry all the scale dependence and are given by (\ref{OPE}), whereas the hadronic form factor $f_{\rm hadr}(r)$ with $f_{\rm hadr}(0)=1$ describes the fall-off at large transverse distances and is parameterized in terms of a hadronic scale $\Lambda_{\rm NP}$. For simplicity, we will assume that this form factor is independent of $\xi$. The above ansatz inserts a factor $[f_{\rm hadr}(x_T\Lambda_{\rm NP})]^2$ under the integral over $x_T$ in (\ref{Cfinal}), which suppresses the region of very large $x_T$ values. We will employ the Gaussian model
\begin{equation}\label{fmodels}
   f_{\rm hadr}^{\rm Gauss}(x_T\Lambda_{\rm NP}) = \exp\left( -\Lambda_{\rm NP}^2\,x_T^2 \right)
\end{equation}
for the form factor. For the case of Drell-Yan production, it was shown in \cite{Becher:2011xn} that the functional form of the model function only has a minor impact on the results, which are mainly sensitive to the value of the parameter $\Lambda_{\rm NP}$, and we have confirmed that the same is true in the present case. Choosing $\Lambda_{\rm NP}\approx 600$\,MeV shifts the position of the peak of the $q_T$ distribution for $Z$-boson production at the LHC from 3.2\,GeV to 3.5\,GeV and yields to a significantly better agreement with the data. A similar effect is seen for Tevatron data.

\begin{figure}[t!]
\begin{center}
\begin{tabular}{cc}
\includegraphics[width=0.445\textwidth]{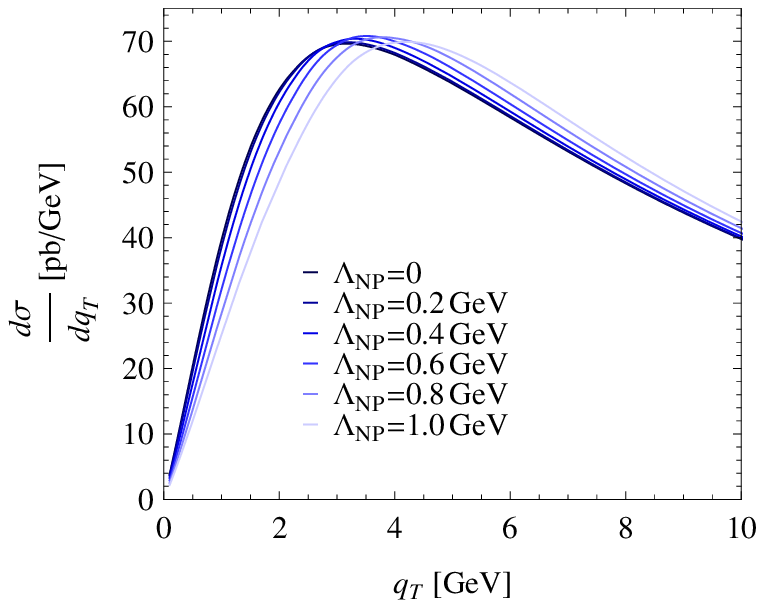} &
\includegraphics[width=0.45\textwidth]{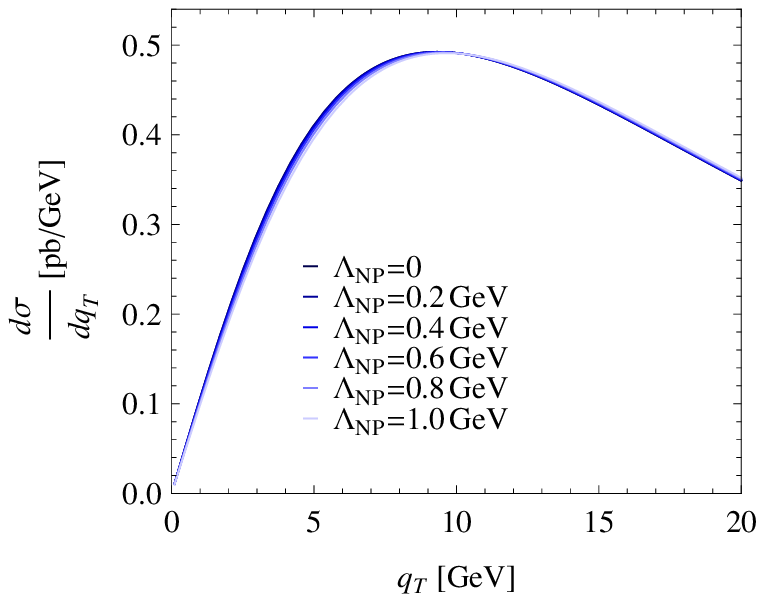}
\end{tabular}
\vspace{-0.5cm}
\end{center}
\caption{\label{fig:nonpert}
Comparison of the importance of long-distance hadronic effects on the differential cross sections $d\sigma/dq_T$ for $Z$-boson (left) and Higgs-boson production (right) at the LHC with $\sqrt{s}=8$\,TeV. We adopt the Gaussian model (\ref{fmodels}) and vary $\Lambda_{\rm NP}$ between 0 and 1\,GeV. The cross sections for $Z$-boson production include a factor of ${\rm Br}(Z\to\ell^+\ell^-)=3.37\%$.}
\end{figure}

In Figure~\ref{fig:nonpert}, we compare the situation in Drell-Yan production of $Z$ bosons, for which the characteristic scale $q_*\approx 1.75$\,GeV is rather low, with that in Higgs production at the LHC, for which $q_*\approx 7.7$\,GeV is safely in the perturbative domain. As expected, we find that the impact of hadronic effects is significantly reduced in the latter case. With $\Lambda_{\rm NP}\approx 600$\,MeV, for instance, the peak position shifts by merely 100\,MeV (from 9.1\,GeV to 9.2\,GeV), which is hardly visible on the scale of the plot. Even for $\Lambda_{\rm NP}=1\,{\rm GeV}$, the shift amounts to only 300\,MeV. We will see in the next section that perturbative uncertainties are significantly larger than this effect. As long as $\Lambda_{\rm NP}$ is in the GeV range, it therefore seems safe to ignore the potential impact of long-distance effects for all practical purposes.

\section{Predictions for the LHC}
\label{sec:LHC}

Having discussed the factorization of the cross section and its behavior at very small $q_T$, we now present our final results for the transverse-momentum spectrum of Higgs bosons produced in gluon fusion at the LHC. In order to obtain reliable predictions also at intermediate $q_T$ values, we match the resummed differential cross section (\ref{fact2}) to the ${\cal O}(\alpha_s)$ fixed-order result. To this end, we add the fixed-order cross section $\sigma_{\rm NLO}$ to the resummed result and subtract the fixed-order expansion of (\ref{fact2}) so as to avoid double counting:
\begin{equation}
   d\sigma_{\rm NNLL+NLO} = d\sigma_{\rm NNLL} + d\sigma_{\rm matching} 
   = d\sigma_{\rm NNLL} + \big( d\sigma_{\rm NLO} - d\sigma_{\rm NNLL}\big|_{\text{expanded to NLO}} 
    \big) \,.
\end{equation}
The expansion of the resummed result to ${\cal O}(\alpha_s)$ can be derived using
\begin{equation}\label{myresult}
\begin{aligned}
   q_T^2\,\bar C_{gg\to ij}(z_1,z_2,q_T^2,m_H^2,\mu) 
   &= \frac{a_s}{2}\,\bigg[ \left( 2\Gamma_0^A\,\ln\frac{q_T^2}{m_H^2} + 4\gamma_0^g \right) 
    \delta(1-z_1)\,\delta(1-z_2)\,\delta_{gi}\,\delta_{gj} \\ 
   &\hspace{1.3cm}\mbox{}+ {\cal P}_{g\leftarrow i}^{(1)}(z_1)\,\delta(1-z_2)\,\delta_{gj} 
    + \delta(1-z_1)\,\delta_{gi}\,{\cal P}_{g\leftarrow j}^{(1)}(z_2) \bigg] \,.
\end{aligned}
\end{equation}
Since the hard function $H(\mu)=C_t^2(m_t^2,\mu)\left|C_S(-m_H^2,\mu)\right|^2=1+{\cal O}(\alpha_s)$, the corrections encoded in $H(\mu)$ do not contribute to the ${\cal O}(\alpha_s)$ expansion of the resummed cross section. Nevertheless, from a physical point of view one may argue that one should factor out these corrections in (\ref{myresult}), because they are to a large extent universal. This is obvious for the prefactor $C_t^2$, but it should also be the case for the large Sudakov logarithms and other corrections encoded in $|C_S|^2$, which also appear in the total cross section \cite{Ahrens:2008qu,Ahrens:2008nc}. Factoring out the hard function $H(\mu)$ before performing the matching leads to a Sudakov suppression of the matching correction at low $q_T$ and an enhancement at larger transverse momentum. 

\begin{figure}[t!]
\begin{center}
\begin{tabular}{ccc}
\hspace{-0.3cm}\includegraphics[height=0.283\textwidth]{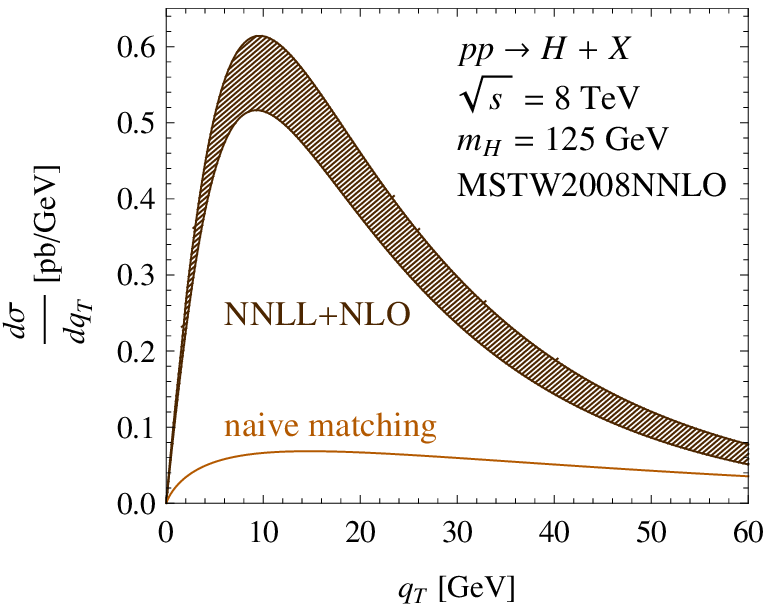}\hspace{-0.6cm} & \includegraphics[height=0.283\textwidth]{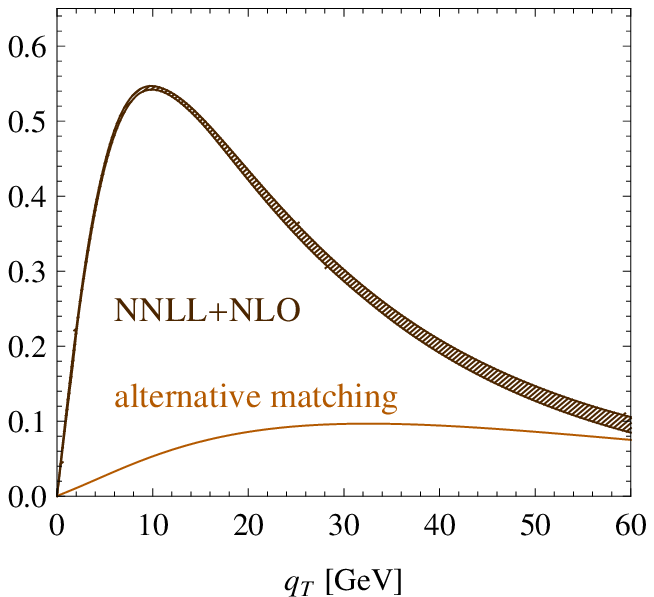} & \hspace{-0.7cm}
\includegraphics[height=0.283\textwidth]{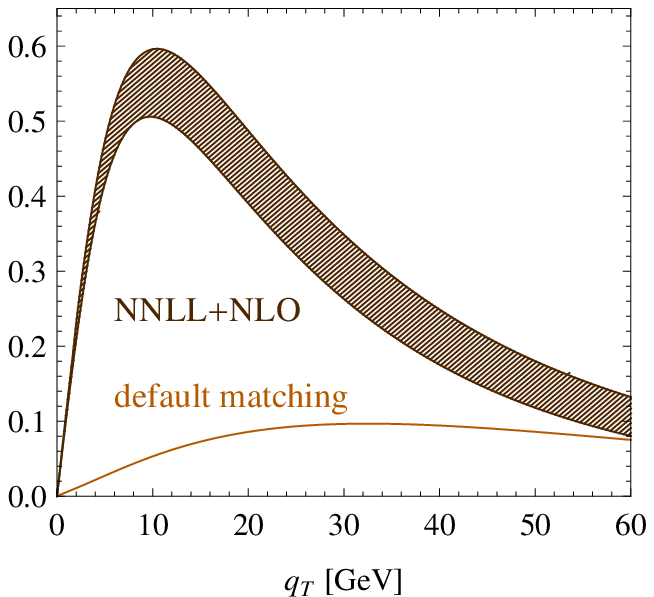} 
\end{tabular}
\end{center}
\vspace{-0.5cm}
\caption{\label{fig:matching}
Comparison of different schemes for performing the matching to fixed-order results at ${\cal O}(\alpha_s)$. The bands show the resummed and matched results, the line the matching correction itself.}
\end{figure}

In Figure~\ref{fig:matching}, we compare our results for the resummed and matched differential cross section obtained when the hard function is not factored out (``naive matching'') with the those found when the right-hand side of (\ref{myresult}) is multiplied by $H(\mu)$ (``alternative matching''). In the second scheme, we indeed observe the Sudakov suppression of the matching correction at low $q_T$, but also a drastic reduction of the scale dependence as displayed by the width of the band. This is due to a cancellation of the $\mu$ dependence between the resummed result and the matching correction. While some reduction may be expected, such a strong effect is presumably in part accidental. Indeed, instead of using $H(\mu)$ with variable $\mu$ we may equally well factor out the hard function at the fixed default scale $\mu=q_T+q_*$. Doing so has the same qualitative effect on the matching correction, but as shown in the third plot the strong cancellation of scale dependence is not observed. To be conservative, we adopt this last choice as our ``default matching'' prescription. 

Since the matching correction for Higgs-boson production is several times larger than that for the Drell-Yan case (see e.g.\ \cite{Becher:2011xn}), it would be preferable to extend the matching to the fixed-order cross section to two-loop order. This requires some effort, but it is possible since the corresponding fixed-order result is known \cite{deFlorian:1999zd,Ravindran:2002dc,Glosser:2002gm} and has been implemented in several public codes, e.g.\ MCFM \cite{mcfm} and HNNLO \cite{Grazzini:2008tf}. We note that the quark beam function $I_{q\to q}(z,x_T^2,\mu)$ has recently been computed to two-loop accuracy \cite{Gehrmann:2012ze}. Once this result is extended to the gluon channel, all two-loop ingredients for the resummed expression (\ref{fact2}) will be known, and the matching should then be extended to ${\cal O}(\alpha_s^2)$. 

\begin{figure}[t!]
\begin{center}
\begin{tabular}{rcr}
\includegraphics[width=0.45\textwidth]{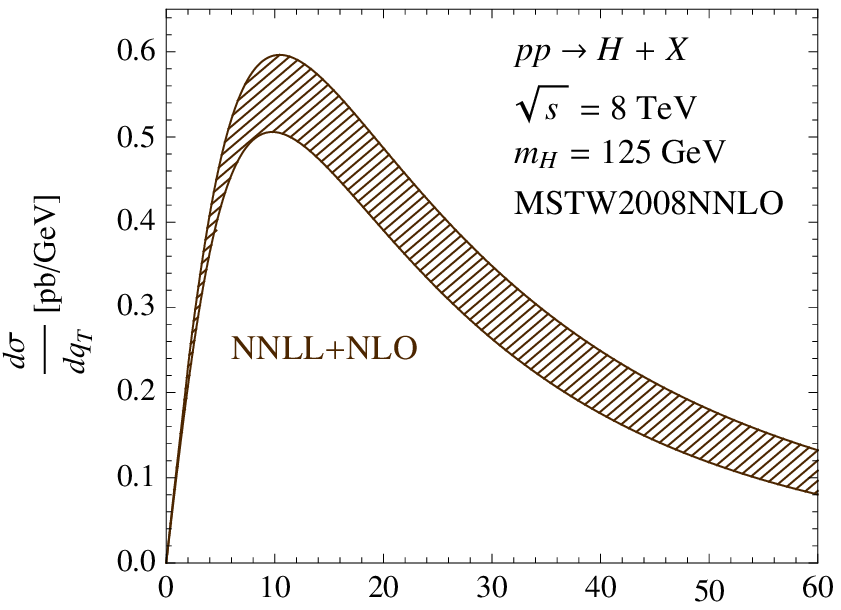} &
 & \includegraphics[width=0.45\textwidth]{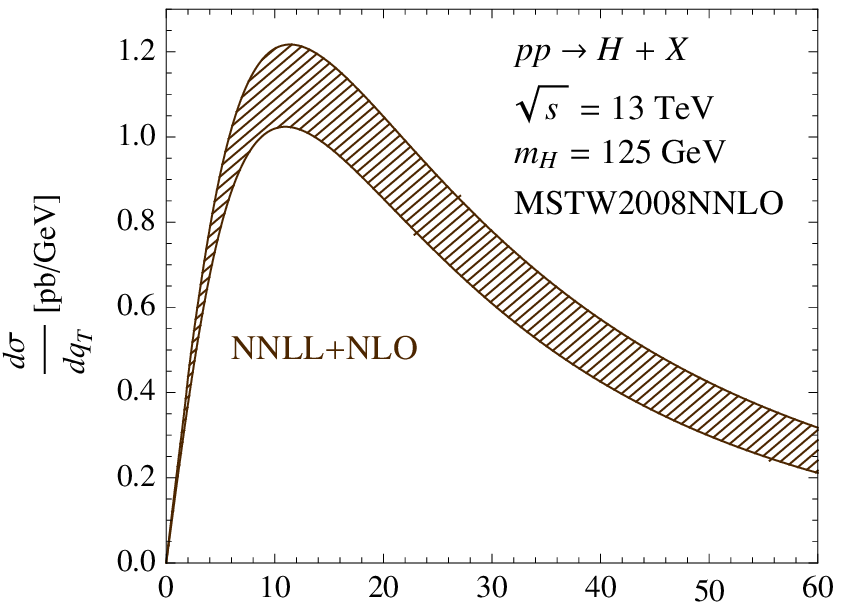} \\[-0.5cm]
 \includegraphics[width=0.42\textwidth]{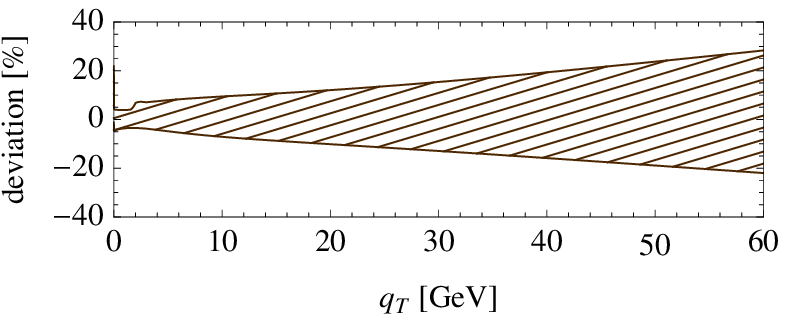} &
 & \includegraphics[width=0.42\textwidth]{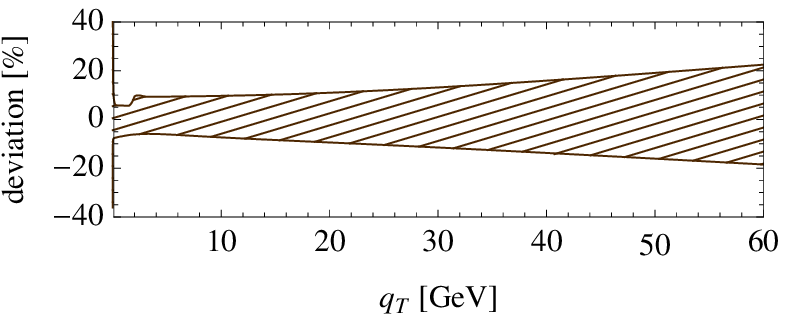} 
\end{tabular}
\end{center}
\vspace{-0.5cm}
\caption{\label{fig:matching_final}
Comparison of the resummed and matched transverse-momentum distributions of Higgs bosons produced at the LHC, for $\sqrt{s}=8$\,TeV (left) and $\sqrt{s}=13$\,TeV (right). The default matching scheme is adopted. Note the different scales in the plots.}
\end{figure}

Our final results for the resummed and matched differential cross sections for Higgs production at the LHC, for $\sqrt{s}=8$\,TeV and 13\,TeV, are shown in Figure~\ref{fig:matching_final}. The shape of the two spectra is very similar, the main effect of the higher center-of-mass energy being an increase in the cross section by about a factor of 2. The scale uncertainty is around $\pm 10\%$ in the peak region and increases for larger $q_T$, as indicated in the panels below the plots. Our results are fully compatible with the NNLL order predictions of \cite{deFlorian:2011xf} obtained in the traditional CSS resummation framework \cite{Collins:1984kg}. The uncertainties found in that paper are slightly smaller in the peak region, but about a factor of 2 smaller at large $q_T$. The reason for the reduced scale uncertainty is that this work implements matching to ${\cal O}(\alpha_s^2)$ as well as the hard-collinear two-loop corrections, which were calculated in \cite{Catani:2011kr}. 

\begin{figure}[t!]
\begin{center}
\begin{tabular}{r}
\includegraphics[width=0.45\textwidth]{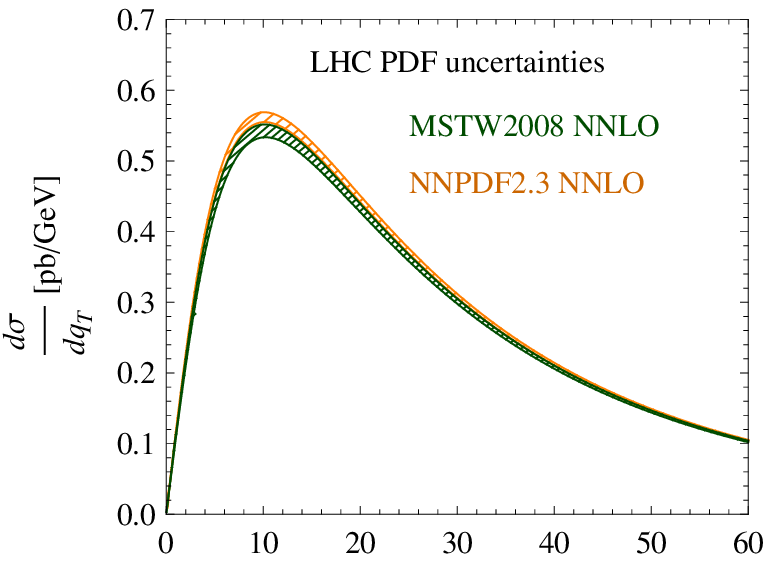} \\[-0.5cm]
\includegraphics[width=0.42\textwidth]{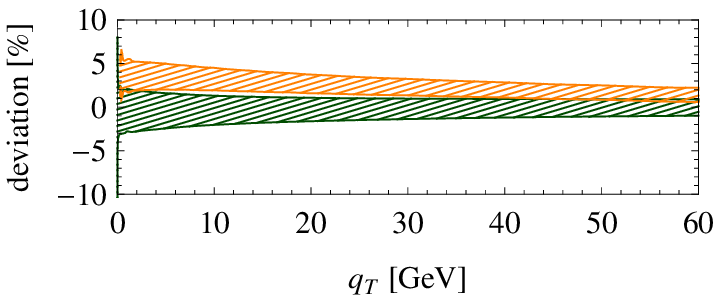} 
\end{tabular}
\end{center}
\vspace{-0.5cm}
\caption{\label{pdfuncertainty}
Estimate of PDF uncertainties for Higgs production at the LHC with $\sqrt{s}=8$\,TeV. The bands  correspond to one standard deviation variations of the PDF sets as compiled in MSTW2008~NNLO (green) and NNPDF~2.3 (orange). They do not include the uncertainty associated with the value of $\alpha_s$. For NNPDF, we choose the PDF set with the MSTW default value $\alpha_s(M_Z)=0.1170$.}
\end{figure}

We have implemented the results of our calculations into a public numerical code {\tt CuTe} \cite{CuTe}, which produces resummed and matched NNLL+NLO results not only for Higgs production, but also for the Drell-Yan process and the production of $Z$ and $W$ bosons at hadron colliders. The {\tt CuTe}  program computes cross sections with scale uncertainties and also includes different models for non-perturbative effects. It uses the LHAPDF interface, so it is straightforward to switch between different PDF sets. As an example, we show in Figure~\ref{pdfuncertainty} the result obtained with NNPDF~2.3 PDF sets \cite{Ball:2012cx} and compare it with our default choice. The width of the bands now reflects the PDF uncertainties at the level of one standard deviation. We find good agreement of the distributions obtained with NNPDF~2.3 and MSTW2008~NNLO, the former giving rise to slightly higher cross sections.

\section{Conclusions}

We have extended our previously developed formalism for the calculation of Drell-Yan cross sections at low and very low transverse momentum $q_T$ \cite{Becher:2011xn} to case of Higgs-boson production in gluon fusion at a hadron collider. Large Sudakov double logarithms in $m_H/q_T$ are resummed in a systematic way using RG and anomaly equations. The leading logarithms are contained in the hard matching coefficient $C_S$, which arises when the effective $ggH$ operator is matched onto a two-gluon current operator in SCET. They are resummed by evolving this coefficient from a hard matching scale $\mu_h^2\sim-m_H^2$ down to a characteristic scale $\mu\sim q_T$. At NLL order, additional large logarithms are generated by the collinear factorization anomaly \cite{Becher:2010tm}. In transverse-position space, they give rise to an extra factor of the form $(x_T^2 m_H^2)^{-F_{gg}(x_T^2,\mu)}$. Long-distance hadronic effects are controlled by a dynamically generated scale $q_*$, which sets an upper limit on the average transverse separation in the process, $\langle x_T\rangle\lesssim 1/q_*$. As long as the value of $q_*$ lies in the perturbative domain, the transverse-momentum distribution is short-distance dominated even for very small values of $q_T$. Due to the difference in color factors ($C_A$ vs.\ $C_F$), this scale is significantly larger in the case of Higgs production ($q_*\approx 7.7$\,GeV) compared with that of $Z$-boson production ($q_*\approx 1.75$\,GeV). As a result, we find that long-distance hadronic corrections to the shape of the transverse-momentum distribution of Higgs bosons produced at the LHC have a negligible effect, see Figure~\ref{fig:nonpert}.

From a technical point of view, a novel feature of the case of Higgs production is that the tensor structure of the beam-jet functions $B_g^{\mu\nu}$ in (\ref{Bmunudec}) implies that the factorized cross sections (\ref{sig3}) and (\ref{fact1}) cannot be expressed as a simple product of two beam functions, but rather as a sum over products, reflecting the entanglement of the spin-0 state of two gluons \cite{Catani:2010pd}. However, we have shown that one can find a basis in which this effect arises first at NNLO in $\alpha_s$.

Our numerical analysis of the transverse-momentum distribution of Higgs bosons produced in gluon fusion at the LHC (with $\sqrt{s}=8$ and 13\,TeV) presented in Section~\ref{sec:LHC} gives rise to robust predictions, which hopefully will soon be confronted with first data from the LHC. We have explored different schemes for performing the matching to fixed-order perturbation theory and found that the various results agree within the estimated theoretical uncertainties. Our final predictions are presented in Figure~\ref{fig:matching_final}, with PDF uncertainties estimated in Figure~\ref{pdfuncertainty}. They suggest that at NNLL+NLO our predictions for the differential cross section $d\sigma/dq_T$ have uncertainties of order $\pm 10\%$ in the peak region, which seems reasonable given what is known about the behavior of the perturbative series for the total cross section. In order to reduce these uncertainties further, it would be necessary and desirable to extend our analysis to N$^3$LL+NNLO, or at least to perform the matching to fixed-order theory at two-loop order. 

\vspace{0.2cm}
{\em Acknowledgments:\/}
The work of T.B.\ is supported by the Swiss National Science Foundation (SNF) under grant 200020-140978 and the Innovations- und Kooperationsprojekt C-13 of the Schweizerische Universit\"atskonferenz (SUK/CRUS). The research of M.N.\ and D.W.\ is supported by the ERC Advanced Grant EFT4LHC of the European Research Council, the Cluster of Excellence {\em Precision Physics, Fundamental Interactions and Structure of Matter\/} (PRISMA -- EXC 1098) and grant NE 398/3-1 of the German Research Foundation (DFG), grants 05H09UME and 05H12UME of the German Federal Ministry for Education and Research (BMBF), and the Rhineland-Palatinate Research Center {\em Elementary Forces and Mathematical Foundations}.



\begin{thebibliography}{99}

\bibitem{Collins:1984kg}
  J.~C.~Collins, D.~E.~Soper and G.~F.~Sterman,
  Nucl.\ Phys.\  B {\bf 250}, 199 (1985).

\bibitem{Ladinsky:1993zn}
  G.~A.~Ladinsky and C.~P.~Yuan,
  Phys.\ Rev.\  D {\bf 50}, 4239 (1994)
  [arXiv:hep-ph/9311341].
  
\bibitem{Cao:2009md} 
  Q.~-H.~Cao, C.~-R.~Chen, C.~Schmidt and C.~-P.~Yuan,
  arXiv:0909.2305 [hep-ph].

\bibitem{deFlorian:2011xf} 
  D.~de Florian, G.~Ferrera, M.~Grazzini and D.~Tommasini,
  JHEP {\bf 1111}, 064 (2011)
  [arXiv:1109.2109 [hep-ph]].

\bibitem{Bozzi:2010xn}
  G.~Bozzi, S.~Catani, G.~Ferrera, D.~de Florian and M.~Grazzini,
  Phys.\ Lett.\  B {\bf 696}, 207 (2011)
  [arXiv:1007.2351 [hep-ph]].
  
\bibitem{Wang:2012xs} 
  J.~Wang, C.~S.~Li, Z.~Li, C.~P.~Yuan and H.~T.~Li,
ÊÊPhys.\ Rev.\ D {\bf 86}, 094026 (2012)
ÊÊ[arXiv:1205.4311 [hep-ph]].
ÊÊ

\bibitem{Catani:2011kr} 
  S.~Catani and M.~Grazzini,
  Eur.\ Phys.\ J.\ C {\bf 72}, 2013 (2012)
  [Erratum-ibid.\ C {\bf 72}, 2132 (2012)]
  [arXiv:1106.4652 [hep-ph]].
  
  \bibitem{Bauer:2000yr}
 C.~W.~Bauer, S.~Fleming, D.~Pirjol and I.~W.~Stewart,
 Phys.\ Rev.\  D {\bf 63}, 114020 (2001)
 [arXiv:hep-ph/0011336].

\bibitem{Bauer:2001yt}
 C.~W.~Bauer, D.~Pirjol and I.~W.~Stewart,
 Phys.\ Rev.\  D {\bf 65}, 054022 (2002)
 [arXiv:hep-ph/0109045].

\bibitem{Beneke:2002ph}
 M.~Beneke, A.~P.~Chapovsky, M.~Diehl and T.~Feldmann,
 Nucl.\ Phys.\  B {\bf 643}, 431 (2002)
 [arXiv:hep-ph/0206152].
  
\bibitem{Becher:2010tm}
  T.~Becher and M.~Neubert,
  Eur.\ Phys.\ J.\  C {\bf 71}, 1665 (2011)
  [arXiv:1007.4005 [hep-ph]].

\bibitem{Chiu:2012ir} 
  J.~-Y.~Chiu, A.~Jain, D.~Neill and I.~Z.~Rothstein,
  JHEP {\bf 1205}, 084 (2012)
  [arXiv:1202.0814 [hep-ph]].
  
\bibitem{GarciaEchevarria:2011rb} 
  M.~G.~Echevarria, A.~Idilbi and I.~Scimemi,
  JHEP {\bf 1207}, 002 (2012)
  [arXiv:1111.4996 [hep-ph]].

\bibitem{Gao:2005iu}
 Y.~Gao, C.~S.~Li and J.~J.~Liu,
 Phys.\ Rev.\  D {\bf 72}, 114020 (2005)
 [arXiv:hep-ph/0501229].

\bibitem{Idilbi:2005er}
 A.~Idilbi, X.~d.~Ji and F.~Yuan,
 Phys.\ Lett.\  B {\bf 625}, 253 (2005)
 [arXiv:hep-ph/0507196].
  
\bibitem{Mantry:2009qz}
 S.~Mantry and F.~Petriello,
 Phys.\ Rev.\  D {\bf 81}, 093007 (2010)
 [arXiv:0911.4135 [hep-ph]].
   
\bibitem{Becher:2011xn} 
  T.~Becher, M.~Neubert and D.~Wilhelm,
  JHEP {\bf 1202}, 124 (2012)
  [arXiv:1109.6027 [hep-ph]].

\bibitem{Parisi:1979se}
  G.~Parisi and R.~Petronzio,
  Nucl.\ Phys.\  B {\bf 154}, 427 (1979).

\bibitem{Banfi:2012yh} 
  A.~Banfi, G.~P.~Salam and G.~Zanderighi,
  JHEP {\bf 1206}, 159 (2012)
  [arXiv:1203.5773 [hep-ph]].

\bibitem{Becher:2012qa} 
  T.~Becher and M.~Neubert,
  JHEP {\bf 1207}, 108 (2012)
  [arXiv:1205.3806 [hep-ph]].

\bibitem{Tackmann:2012bt} 
  F.~J.~Tackmann, J.~R.~Walsh and S.~Zuberi,
  Phys.\ Rev.\ D {\bf 86}, 053011 (2012)
  [arXiv:1206.4312 [hep-ph]].

\bibitem{Banfi:2012jm} 
  A.~Banfi, P.~F.~Monni, G.~P.~Salam and G.~Zanderighi,
  Phys.\ Rev.\ Lett.\  {\bf 109}, 202001 (2012)
  [arXiv:1206.4998 [hep-ph]].


\bibitem{Liu:2012sz} 
  X.~Liu and F.~Petriello,
  Phys.\ Rev.\ D {\bf 87}, 014018 (2013)
  [arXiv:1210.1906 [hep-ph]].


\bibitem{Catani:2010pd} 
  S.~Catani and M.~Grazzini,
  Nucl.\ Phys.\ B {\bf 845}, 297 (2011)
  [arXiv:1011.3918 [hep-ph]].

\bibitem{CuTe}
  T.~Becher, M.~Neubert and D.~Wilhelm, http://cute.hepforge.org.

\bibitem{Collins:1981uw} 
  J.~C.~Collins and D.~E.~Soper,
  Nucl.\ Phys.\ B {\bf 194}, 445 (1982).


\bibitem{Stewart:2009yx} 
  I.~W.~Stewart, F.~J.~Tackmann and W.~J.~Waalewijn,
ÊÊPhys.\ Rev.\ D {\bf 81}, 094035 (2010)
ÊÊ[arXiv:0910.0467 [hep-ph]].
ÊÊ

\bibitem{Stewart:2010qs} 
  I.~W.~Stewart, F.~J.~Tackmann and W.~J.~Waalewijn,
  JHEP {\bf 1009}, 005 (2010)
  [arXiv:1002.2213 [hep-ph]].
  
\bibitem{Jain:2011iu} 
  A.~Jain, M.~Procura and W.~J.~Waalewijn,
  JHEP {\bf 1204}, 132 (2012)
  [arXiv:1110.0839 [hep-ph]].

\bibitem{Becher:2011dz} 
  T.~Becher and G.~Bell,
  Phys.\ Lett.\ B {\bf 713}, 41 (2012)
  [arXiv:1112.3907 [hep-ph]].
    
\bibitem{Chiu:2007dg} 
  J.-y.~Chiu, F.~Golf, R.~Kelley and A.~V.~Manohar,
  Phys.\ Rev.\ D {\bf 77}, 053004 (2008)
  [arXiv:0712.0396 [hep-ph]].
  
\bibitem{Becher:2009qa} 
  T.~Becher and M.~Neubert,
  JHEP {\bf 0906}, 081 (2009)
  [arXiv:0903.1126 [hep-ph]].
 
\bibitem{Becher:2009cu} 
  T.~Becher and M.~Neubert,
  Phys.\ Rev.\ Lett.\  {\bf 102}, 162001 (2009)
  [arXiv:0901.0722 [hep-ph]].
  
\bibitem{Frixione:1998dw}
  S.~Frixione, P.~Nason and G.~Ridolfi,
  Nucl.\ Phys.\  B {\bf 542}, 311 (1999)
  [arXiv:hep-ph/9809367].
  
\bibitem{private} J.~-Y.~Chiu, A.~Jain, D.~Neill and I.~Z.~Rothstein, private communication.

\bibitem{Martin:2009iq} 
  A.~D.~Martin, W.~J.~Stirling, R.~S.~Thorne and G.~Watt,
  Eur.\ Phys.\ J.\ C {\bf 63}, 189 (2009)
  [arXiv:0901.0002 [hep-ph]].
  
\bibitem{Ahrens:2008nc}
  V.~Ahrens, T.~Becher, M.~Neubert and L.~L.~Yang,
  Eur.\ Phys.\ J.\  C {\bf 62}, 333 (2009)
  [arXiv:0809.4283 [hep-ph]].
  
\bibitem{Ahrens:2008qu}
  V.~Ahrens, T.~Becher, M.~Neubert and L.~L.~Yang,
  Phys.\ Rev.\  D {\bf 79}, 033013 (2009)
  [arXiv:0808.3008 [hep-ph]].

\bibitem{Konychev:2005iy}
  A.~V.~Konychev and P.~M.~Nadolsky,
  Phys.\ Lett.\ B {\bf 633}, 710 (2006) 
  [arXiv:hep-ph/0506225].

\bibitem{deFlorian:1999zd} 
  D.~de Florian, M.~Grazzini and Z.~Kunszt,
  Phys.\ Rev.\ Lett.\  {\bf 82}, 5209 (1999)
  [hep-ph/9902483].
  
\bibitem{Ravindran:2002dc} 
  V.~Ravindran, J.~Smith and W.~L.~Van Neerven,
  Nucl.\ Phys.\ B {\bf 634}, 247 (2002)
  [hep-ph/0201114].

\bibitem{Glosser:2002gm} 
  C.~J.~Glosser and C.~R.~Schmidt,
  JHEP {\bf 0212}, 016 (2002)
  [hep-ph/0209248].

\bibitem{mcfm}
  J.~Campbell, K.~Ellis and C.~Williams, http://mcfm.fnal.gov/.

\bibitem{Grazzini:2008tf} 
  M.~Grazzini,
  JHEP {\bf 0802}, 043 (2008)
  [arXiv:0801.3232 [hep-ph]].
  
\bibitem{Gehrmann:2012ze} 
  T.~Gehrmann, T.~L\"ubbert and L.~L.~Yang,
%
  Phys.\ Rev.\ Lett.\  {\bf 109}, 242003 (2012)
  [arXiv:1209.0682 [hep-ph]].
%

\bibitem{Ball:2012cx} 
  R.~D.~Ball, V.~Bertone, S.~Carrazza, C.~S.~Deans, L.~Del Debbio, S.~Forte, A.~Guffanti and N.~P.~Hartland {\it et al.},
  Nucl.\ Phys.\ B {\bf 867}, 244 (2013)
  [arXiv:1207.1303 [hep-ph]].

\end{thebibliography}
\end{document}